\documentclass[a4paper, 12pt]{article}

\usepackage[margin=3cm]{geometry}
\usepackage{graphicx}% Include figure files
\usepackage{dcolumn}% Align table columns on decimal point
\usepackage{comment}
\usepackage{amsmath}
\usepackage{subcaption}
\usepackage{amsfonts}
\usepackage{datetime}
\usepackage{xcolor} % optional

\begin{document}

%\small{{\color{red}\textbf{version: \today, \currenttime}}}

\begin{center}
    \begin{large}
            \textbf{Identifying ballistic modes via Poincaré sections}
    \end{large}
    
    \vspace{0.25cm}
    
    A.F. Bósio$^{1,3}$, I.L. Caldas$^1$, R.L. Viana$^2$, Y. Elskens$^3$
    
    \vspace{0.25cm}

\begin{small}
        $^1$ Universidade de São Paulo, R. da Reitoria, 374, Butantã, São Paulo, São Paulo, Brazil 05508220\\ $^2$ Department of Physics, Universidade Federal do Paraná, Centro Interdisciplinar de Ciência, Tecnologia e Inovação, Nùcleo de Modelagem e Computação Científica, Curitiba-PR, Brazil 81530990 \\ $^3$ Aix-Marseille Université, UMR 7345 CNRS, PIIM, case 322 campus Saint-Jérôme, \\
        52, av. Escadrille Normandie-Niemen, Marseille, France 13013 
    \end{small}    
\end{center}

\date{\today}% It is always \today, today,

\begin{abstract}
Exploring chaotic systems via Poincaré sections has proven essential in dynamical systems, yet measuring their characteristics poses challenges to identify the various dynamical regimes considered. In this paper, we propose a new approach that uses image processing to distinguish chaotic and regular regions of area-preserving dynamics, and then classify the transport regime. We characterize different transport regimes in the standard map with the proposed method based on image reconstruction techniques, identifying the superdiffusion much faster than the usual mean square displacement method. The procedure is also applied to a two-wave, time-dependent Hamiltonian to investigate superdiffusion in function of two parameters.


\end{abstract}

%\tableofcontents

\section{\label{sec:level1} Introduction}

Chaotic transport is a topic of paramount interest in the study of conservative, non-integrable dynamical systems. In many systems of physical interest, the observed transport properties often differ from those predicted by classical diffusion \cite{klages2008anomalous}. The chaotic transport that differs from predicted diffusion is called anomalous transport, and may have different sources, such as long-range correlations, memory effects, transport channels, and biased statistics. Anomalous transport can appear in plasmas \cite{balescu2005aspects,wootton1990fluctuations}, biological systems \cite{hofling2013anomalous} and others \cite{klages2008anomalous}, so it is important to characterize this process.

Amongst many sources of anomalous transport, one that is relevant in the scope of this work is the ballistic mode, as its presence guarantees this transport regime, as orbits that pass close to these modes perform long flights \cite{shlesinger1993strange,metzler2014anomalous,klages2008anomalous}. 

A simple and widely used method to numerically identify anomalous transport is to observe the trajectories of a large number $N$ of initial conditions and then see how the ensemble's mean square displacement (MSD) evolves over time. By fitting a power law on the MSD, we get an exponent $\gamma$ that characterizes the transport. If $\gamma < 1$, the transport regime is subdiffusive; if the system behaves according to normal diffusion, then $\gamma = 1$, which is the case for Brownian motion; and if $\gamma > 1$, one has superdiffusion.

For discrete systems such as the standard map (also known as Chirikov-Taylor map) \cite{CHIRIKOV1979263}, the MSD approach is not an issue, as for each time step, one point in the Poincaré section is obtained. But, for a system of ordinary differential equations (ODEs), it can require a large number of intermediate steps to get a single point on a Poincaré sections, as is the case in two-dimensional Hamiltonian flows \cite{horton1985onset} and optical lattices \cite{lazarotto2022diffusion}. 

In this paper, we take advantage of the fact that ballistic and regular islands have the same appearance in Poincaré section that are periodic in one or more coordinates. Given this, we used image processing techniques to differentiate the periodic and chaotic regions using morphology, and with that, look out for regions that could enhance transport, in a fast manner. As a comparison, using the MSD method requires around $10^3$ initial conditions and around $10^4$ iterations for a good convergence of the exponent $\gamma$. With the proposed method, superdiffusion was identified using a rather small number of initial conditions, instead of the large number used in computations based on the MSD, for the same time (measured in number of map iterations). The morphology approach was chosen as it is already optimized, and this kind of tool is widely used both in academic and industrial environments \cite{paul2006characterisation,way2021predicting,mikli2001characterization,kulu1998possibilities,eccher2010image,farinha2022phase}.

This work is organized as follows: In Section II, we present some morphological transformations, the method itself, its features, and limitations. Section III applies the method to the well-known Chirikov-Taylor map. In Section IV, we apply the method to a non-integrable Hamiltonian system. Section V contains some final remarks about the obtained results.

%\cite{paul2006characterisation,way2021predicting}, granulometry \cite{mikli2001characterization,kulu1998possibilities}, and liquid crystal transitions \cite{eccher2010image,farinha2022phase}. 

\section{\label{sec:level1} The method}

To identify and differentiate regions in phase space, we need to segment (i.e.\ partition) the phase space into different parts. Here, this will be done by converting the data from the Poincaré section into a binary image. This image will pass through filterings using morphological operations; after that, the filtered image is segmented. At the core of each segment, there is an initial condition (IC) that is used to iterate the map. Based on the behavior of this orbit, the region that encloses this IC is labeled accordingly. 

The first step is to generate the data of the Poincaré section using some $M$ initial conditions, on an evenly spaced grid. A $9 \times 9$ grid was sufficient in the systems tested in this paper. Each IC is iterated $N$ times, so that we have the resulting set of points, $S = \{s_{10}, s_{11}, s_{12},..., s_{1N}, s_{20}, s_{21}, s_{22},..., s_{MN}\}$. Here, $s_{mn} = (x_{mn},y_{mn})$ is the $n$-th iteration of the $m$-th initial condition.
 
To apply the morphological operations, first, we must transform this set of points $S \subset \mathbb{R}^2$ to a binary image $I \subset \mathbb{Z}^2$. To do so, we select a resolution for our image, say $R_x$ by $R_y$ pixels, and then we define an overlapping lattice of pixels over the phase space, creating the sets
\begin{equation}
    G_{i,j} =  \left\{ (x,y) \in S \Bigg |
    \begin{matrix}
        &i\frac{L_x}{R_x} + l_{x} < x \leq (i+1)\frac{L_x}{R_x}+l_{x} \\
        &j\frac{L_y}{R_y} + l_{y} < y \leq (j+1)\frac{L_y}{R_y}+l_{y}
    \end{matrix}
    \right\}
\end{equation}
with $L_x$ and $L_y$ the side length of the phase space of each coordinate, $l_{x}$ and $l_{y}$ being the lower boundary of the phase space. The indices $i$ and $j$ go from 0 to $R_x - 1$ and $R_y - 1$, respectively. If $G_{ij} \neq \emptyset$, then the pixel $(i, j) \in I$, with $I$ being the image of the Poincaré section. This effectively means that the pixel value $p_{ij}$ is set to 1 (white) if there is at least one point inside the rectangle defined by $G_{ij}$; otherwise, it is set to 0 (black).

Now, we proceed to the filtering process that involves two morphological operations using a structuring element $I_s$, the shape that will be filtered from the image. Usual options are a small cross, disk, or rectangle \cite{gonzalez2009digital}. 

The first filtering is a closing, which fills in any black regions smaller than $I_s$. In the context of this paper, it will fill in the chaotic region. The result is the closed image $I_C$ \cite{gonzalez2009digital}.

The second filtering operation is an \textit{opening by reconstruction}, where small bright regions are removed \cite{gonzalez2009digital}, creating the opened image, $I_O$. The opening by reconstruction differs from the opening, by the fact that it does not distort the image like its simple counterpart, but it is computationally expensive. This operation focuses on invariant curves, that could form nested regions, avoiding redundant study of orbits with the same transport properties. 

\begin{figure}[h]
     \centering
     \begin{subfigure}[b]{0.3\textwidth}
         \centering
         \includegraphics[width=\textwidth]{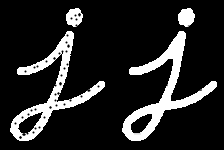}
         \caption{Closing, small black regions removed}
         \label{fig_chaos}
     \end{subfigure}
     \begin{subfigure}[b]{0.3\textwidth}
         \centering
         \includegraphics[width=\textwidth]{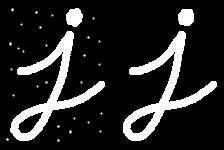}
         \caption{Opening, small white regions removed}
         \label{fig_ball}
     \end{subfigure}
        \caption{Example of the used morphological operations. Extracted from OpenCV documentation \cite{itseez2014theopencv}.}
        \label{fig_operations}
\end{figure}

Figure \ref{fig_operations} displays the general idea behind the morphological filters. In the closing, the black specks inside the white object are filled while maintaining its shape, while in its counterpart, the opening, the white specks are removed from the black background, filtering out the undesirable elements.

With the filtered image $I_O$, we segment (viz.\ partition) the bright and white regions creating the labeled image $I_L$, where each continuous region has its pixels value set to a label $k$, with $k$ ranging from 1 to $K$, $K$ being the number of regions. This way, it is possible to create binary images, or masks, where only pixels with value $k$ are included.

Due to the possibility of regions that are concave or with holes, a simple geometric approach would not work. Then to find the core of each segment, we perform a series of successive erosions by a small structuring element (in this case, a simple cross) until the mask relative to the segment with value $k$ vanishes. Then, we go back one iteration and pick one pixel. 

Following this procedure, each segment has a pixel $(i_k, j_k)$ that is the position of its core, and a direct relation to the phase space, given by
\begin{equation}
    (x_{0k};y_{0k}) = (l_{x} + \frac{L_x}{R_x}i_k;l_{y} + \frac{L_y}{R_y}j_k) \,  . \\
\end{equation}
This initial condition is then iterated according to the dynamical system in question, and its behavior dictates the type of transport present within the whole region, but the specific criterion may be adjusted according to the system, as will be shown.

\section{Application - Standard map}

The Chirikov-Taylor map \cite{CHIRIKOV1979263}, also known as the standard map, 
\begin{equation}
\begin{cases}
    v_{n+1} = v_n + K\sin(\theta_n) \hspace{0.2cm} {\mathrm{mod}} (2 \pi)\\
    \theta_{n+1} = \theta_n + v_{n+1} \hspace{0.2cm} {\mathrm{mod}} (2 \pi) 
\end{cases} 
\label{stdmap}
\end{equation}
is one of the most studied dynamical systems since it presents some interesting phenomena, such as mixed phase space, stickiness, as well as anomalous transport in the momentum $v$. Some literature calls the regular regions where $v$ grows almost linearly as accelerator modes since a steady increase in momentum can be related to an acceleration of the angle $\theta$ \cite{metzler2014anomalous,klages2008anomalous,lichtenberg2013regular,benisti1998nonstandard}. Here we will name islands that display ballistic behavior as ballistic modes.

We repeat our considered method for various values of $0 < K < 9$, using a grid of $9 \times 9$ points evenly distributed in the phase space within the square ($0 < \theta <2\pi$, $-\pi < v <\pi$). Each IC is iterated 50000 times so that the chaotic region is densely filled, to reduce the distortion from the filtering processes. Regarding the parameters for the morphological operations, the chosen resolution is $2048 \times 2048$ pixels for the generated images, so that each pixel corresponds to squares with side $0.0030$ in the phase space. For the filtering process, the structuring element $I_s$ for the closing was a disk with a radius of 3 pixels, which is small compared to the image resolution. For the opening by reconstruction, a disk with a radius of 5 pixels was used.

%In figure \ref{fig_mm_op}, we see how each step changes the image. On the original image, $I$, we notice what we expected to see when observing a mixed phase space. The chaotic region has a granulated aspect, along with the presence of the regular orbits that form rings. In $I_C$, the sea of chaos is filled, and now it is one well-defined white region. In $I_O$, the quasiperiodic orbits were removed, leaving a binary image to be segmented, as shown by $I_L$, where each color represents a different region.

\begin{figure}[h!]
    \centering
    \includegraphics{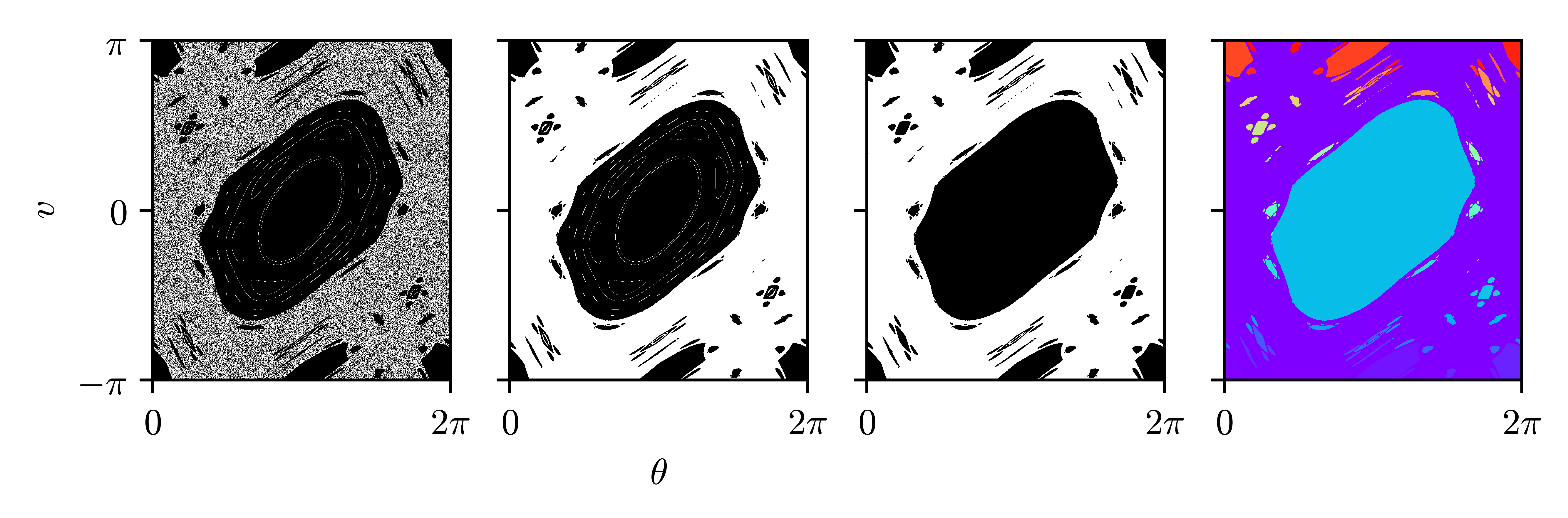}
    \caption{The steps of the routine for the standard map with $K = 1.2625$. From left to right,  original ($I$), closed ($I_C$), opened ($I_O$), segmented regions ($I_{L}$).}
    \label{fig_mm_op}
\end{figure}

Figure \ref{fig_mm_op} displays the morphological steps to obtain the different segments from a the map. $I$ is the original image, obtained by the process described around equation (1), here the chaotic region has a granulated aspect, and within the regular regions the quasiperiodic orbits are noticeable by the thin continuous lines. Following with the filtering, on $I_C$ the chaotic sea was transformed into one continuous white region, the quasiperiodic orbits are intact for now, but disappear during the opening process, resulting on $I_O$, where the quasiperiodic orbits were removed as well, leaving only well defined, continuous black or white regions. In the end, we got the segmented image $I_L$, where each color represents a different segment, or partition element, and each one of them has an IC at its core.

\begin{figure}
    \centering
    \includegraphics{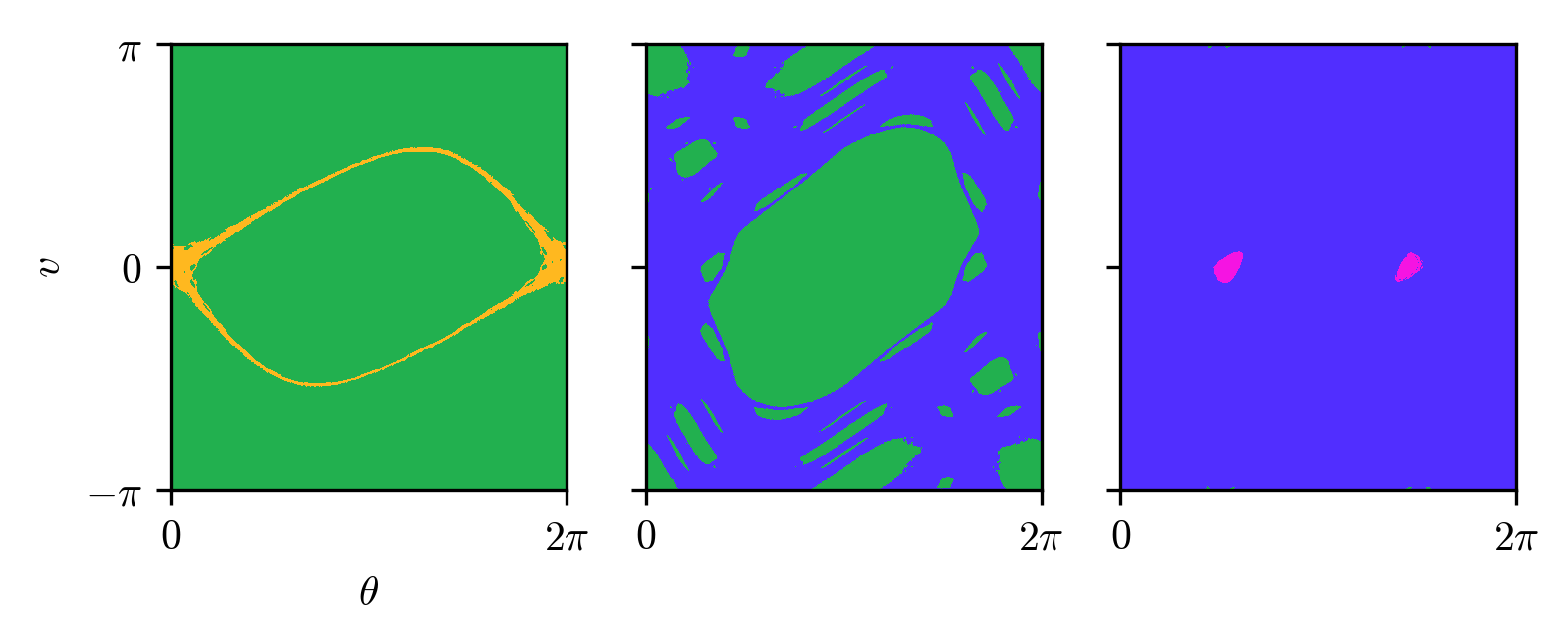}
    \caption{The four region types classified for $K=0.6493$, $1.2084$ and $6.3848$ respectively from left to right. Green pixels represent regular motion, magenta ballistic, yellow and blue bounded and unbounded chaotic respectively.}
    \label{label_type}
\end{figure}

Each region was classified using the orbit from the IC at its core; based on this trajectory, some criteria apply for the classification. First, if the orbit displays ballistic behavior, we label the region as ballistic (B); otherwise, we evaluate the rotation number $\omega(n) = \frac{\theta_i - \theta_0}{n}$ along $\theta$ (disregarding the modulo $2\pi$), if the convergence is good -- in this case, with a standard deviation smaller than $10^{-3}$ --, and the region is labeled as regular (R). Now, the only regimes left are the bounded chaotic (BC) and unbounded chaotic (UC). We check if at any moment the orbit $|\Delta p| > 2\pi$: then it is labeled as UC, and otherwise as BC. These four categories are shown in Figure \ref{label_type}, where each region has a different color representing its regime. For each value of $K$ displayed, each region is classified according to this routine.

\begin{figure}[h!]
    \centering
    \includegraphics{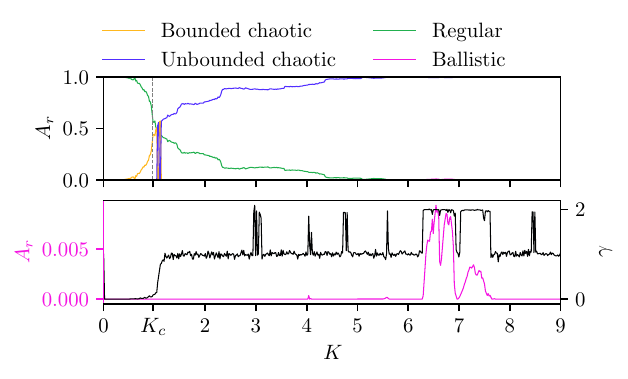}
    \caption{Relative area of each dynamical regime for various values of $K$ (top plot), as well as a comparison between $\gamma$ and the ballistic area, zoomed for better visualization (lower plot).}
    \label{fig_areas}
\end{figure}

Going into the analysis of the method, in Figure \ref{fig_areas}, we compare the relative area (how the total number of pixels of a given regime $A$, relates to the total resolution of the image) $A_r = \frac{A}{R_x \times R_y}$ of each regime against $\gamma$, numerically evaluated via the usual mean square displacement method. The first feature of the plot that needs to be addressed is the oscillation between the bounded and unbounded chaotic areas around $K_c \approx 0.984$. This oscillation occurs very close to the transition to global chaos, and, since the criterion for distinguishing between bounded or unbounded chaos uses a single trajectory, it is subject to fluctuations around transitions like this one. However, it is important to note that this oscillation is restricted to a small range of $K$.

Another feature is that the transition from $\gamma \approx 0$ to $\gamma \approx 1$ for $K \approx 1.0$ is aligned with the sudden rise of the unbounded chaotic regime, showing that the method can detect the transition into large-scale chaos. Although the limit for unbinding $v$ is $K_c = 0.9716$, since the criterion for classification was based on transport properties, some discrepancies are expected, like the displacement of the UC curve. However, at the same time, we have a more consistent result when dealing with the overall transport behavior.

It is also possible to notice abrupt changes in the regular regime, indicating changes in the structure of the phase space. In this case, for $K \approx 1.3$, there is a vanishing of some secondary islands around the main one; and for $K \approx 2.2$, the splitting of the main island into five. Thus, it is possible to identify major changes in the structure of the phase space by looking at the relative areas, rather than examining Poincaré sections with different $K$ one by one.

As it is the main focus of this paper, on the lower plot of Figure \ref{fig_areas}, the magenta line highlights the B curve; since the ballistic modes are small, a magnification is needed. Now, we can compare and see that indeed, when the curve B is nonzero, we always have anomalous transport, as noted for $K \approx 4.050$, $K \approx 5.26$, $6.312 < K < 6.961$ and $ 6.998 < K < 7.611$. As for the regions with $\gamma \approx 2$ that the method did not detect, there could be two reasons. One is simply that the ballistic modes are either smaller than the pixels themselves, or they are smaller than the structuring element of the filtering operations. Since the filtering operations were done using a disk with a radius of 3 pixels for the closing, any island smaller than this size would be lost in the filtering.

With the standard map used as a benchmark, we now move toward a more appropriate use case, which is a system with a non-integrable Hamiltonian.

\section{Continuous time - Electrostatic waves}

One of the major concerns about the current state of plasma confinement is the loss of  particles at the plasma edge \cite{wesson2011tokamaks}. Among many mechanisms, one that displays major importance is particle transport due to drift waves, instabilities that arise from the large pressure gradient at the plasma edge that generate electrostatic instabilities and propagating waves \cite{horton1999drift,horton1985onset}. 

In this context, one model that can give insights into some transport mechanisms was formulated using drift waves \cite{horton1985onset}, where the motion of the guiding centers of charged particles due to $\vec{E} \times \vec{B}$ appears if there is some electric potential $\phi$, constant along the direction of $\vec{B} = B \hat{e_z}$. Then, the equations of motion have a Hamiltonian structure, namely
\begin{equation}
    \vec{v_E} = \frac{\vec{E}\times \vec{B}}{B^2} = \frac{-\nabla \phi \times \vec{B}}{B^2} = 
    -\frac{1}{B}\frac{\partial \phi}{\partial y} \hat{x} + \frac{1}{B}\frac{\partial\phi}{\partial x}  \hat{y} = -\frac{\partial H }{\partial y}\hat{x}  + \frac{\partial H}{\partial x} \hat{y}
\end{equation}
with $y$ being the canonical coordinate and $x$ its conjugate momentum. For the particular case of two drift waves with a static plasma potential, we get the  Hamiltonian 
\begin{equation}
    H(x,y,t) = \frac{\phi_0(x)}{B} + \frac{A_1}{B} \sin(k_{x1}x)\cos(k_{y1}y - \omega_1 t) + \frac{A_2}{B} \sin(k_{x2}x + \theta_x)\cos(k_{y2}y - \omega_2 t)  .
\end{equation}
The constants $B$, $A_i$, $k_{xi}$, $k_{yi}$, $\omega_i$ represent the magnetic field, amplitude of the waves, wavenumbers, and frequency, respectively, and $\phi_0(x)$ the plasma potential. $x$ and $y$ are the radial and poloidal direction of a torus, as is the case for plasmas in tokamaks \cite{horton1985onset}. Since we are interested just in the plasma edge, we can consider the phase space of interest small compared with the whole torus, so that a slab approximation is valid, making the system periodic in $x$ and $y$ \cite{marcus2008reduction}.
  
After a change to a reference frame with the phase velocity of the first wave, $u_1$, and looking into the resonant case where the drift velocity $v_E = \frac{1}{B}\frac{d\phi_0(x)}{dx} = u_1 = \frac{\omega_1}{k_{y1}}$, we obtain the Hamiltonian
\begin{equation}
    H(x,y,t) = \frac{A_1}{B} \sin(k_{x1}x)\cos(k_{y1}y) + \frac{A_2}{B} \sin(k_{x2}x + \theta_x)\cos(k_{y2}(y - u)t)
\end{equation}
where $u = u_2 - u_1 = \frac{\omega_2}{k_{y2}} - \frac{\omega_1}{k_{y1}}$.

When $A_2 = 0$, the system is integrable with no net transport, and the guiding center of each particle remains constrained to curves of constant $H$ around elliptic points, giving rise to a lattice of vortices that closely relates to Taylor-Green vortices \cite{taylor1937mechanism} and geophysical flows \cite{bouchet2012statistical}. When integrability is broken, that is $u \neq 0$, the chaotic region appears along the broken separatrices, which can lead to transport \cite{horton1985onset,marcus2008reduction}. Since there are so many parameters to choose from, such as wavenumbers, phase, and amplitudes, we only highlight some special cases where transport is inhibited as a consequence of the selected parameters.

For convenience, we use integer wavenumbers, so that the phase space always has dimensions $L_x \times L_y = 2\pi \times 2\pi$ and is periodic in both $x$ and $y$. But this also leads to the possibility of transport barriers appearing due to the chosen parameters, particularly about $\theta_x$, $k_{x1}$ and $k_{x2}$.

If the relation
\begin{equation}
    m - n\frac{k_{x2}}{k_{x1}} = \frac{\theta_x}{\pi} \hspace{0.5cm} (m,n \in \mathbb{Z})
\end{equation}
is satisfied, a transport barrier is present and there is no net transport in the $x$ direction \cite{kleva1984stochastic}.

In this work, for the sake of simplicity, we chose $k_{x1} = k_{x2} = k_{y1} = k_{y2} = k = 3$. This way, as long as $\theta_x \neq 0,\pi,2\pi,...$, global transport is guaranteed, leaving us with the Hamiltonian
\begin{equation}
    H(x,y,t) = A_1 \sin(kx)\cos(ky) + A_2 \sin(kx + \theta_x)\cos(k(y - ut)).
\label{2waveh2}
\end{equation}

Anomalous transport has already been reported due to long flights along the broken separatrices as a result of $A_1$ being large \cite{kleva1984stochastic}, so the particles fly along the transport channels that appear as $t$ evolves. 

The source of superdiffusion reported here is due to the presence of ballistic-like modes where for each period $\tau = \frac{2\pi}{ku}$, particles are displaced a somewhat constant value. These sources of superdiffusion are very sensitive to the parameters of the system, and for $k = 3$, $A_1 = 1$ (chosen for convenience), only a small set of combinations of $A_2$ and $\theta_x$ generate this transport behavior.

Since this system is non-integrable, the process of generating a parameter space of $A_2 \times \theta_x$, and evaluating $\gamma$ for each combination of $A_2$ and $\theta_x$, is computationally expensive, making the tool proposed in this paper very useful. Regarding the parameters of the routine, a $9 \times 9$ grid of evenly spaced points was picked from the phase space, each of which was iterated 10000 times to generate the data for the Poincaré sections. For the morphological steps, the resolution of the image was $1024 \times 1024$; for the simple closing operations, a disk of radius 3 was used; and for the opening by reconstruction, a disk of radius 5 was used. 

For the labeling of regions, there are three possible regimes: regular, where no transport occurs; chaotic, which appears along the broken separatrices; and ballistic, which displays ballistic behavior after each time step $\tau$. If the orbit tested does not go as far as $2\sqrt{2}\pi$ at any moment during the integration time, it is confined. If it displays a ballistic behavior, then it is a ballistic regime, and if neither of these applies, by exclusion the behavior must be chaotic. This step also demonstrates the flexibility of the method, as some decision trees are much simpler than others.

\begin{figure}[h]
    \centering
    \includegraphics{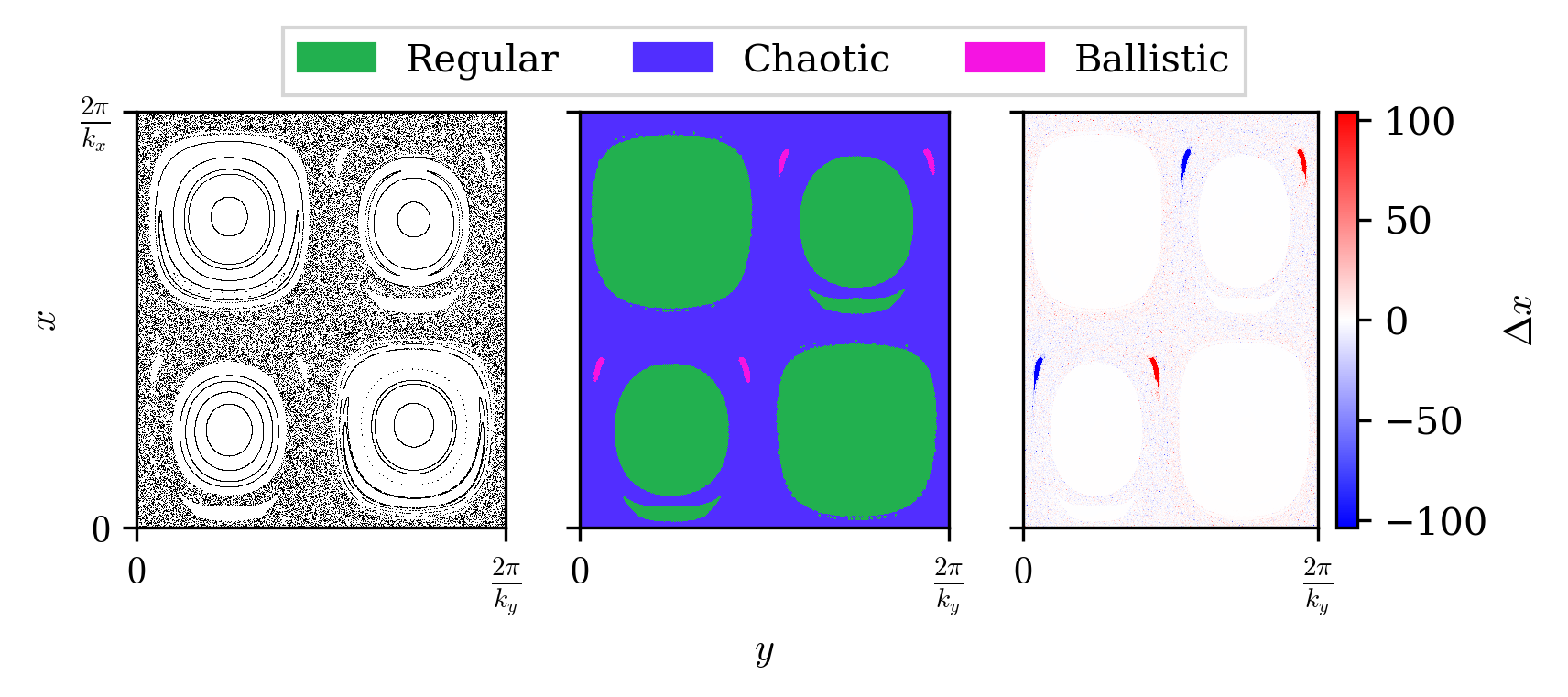}
    \caption{Relation between the Poincaré section (left), labeled regions with the transport type (middle), and displacement from the initial condition on the $x$ direction (left).}
    \label{fig_2w}
\end{figure}

In Figure \ref{fig_2w}, we see a direct comparison between the phase portraits, the classified regions using the proposed method, and the displacement $\Delta x$ relative to each initial condition on the phase space. Comparing the phase portrait with the labeled regions, it is clear that the segmentation was a success, where very little was lost during the filtering process. Again, comparing the labeled regions to the displacement plot, it is also clear that each region was labeled correctly, as the islands with ballistic modes have much higher $|\Delta x|$, the chaotic region still has the granulated aspect characteristic of this behavior, and the regular regions present only a very small displacement.

\begin{figure}[h]
     \centering
     \begin{subfigure}[b]{0.4\textwidth}
         \centering
         \includegraphics[width=\textwidth]{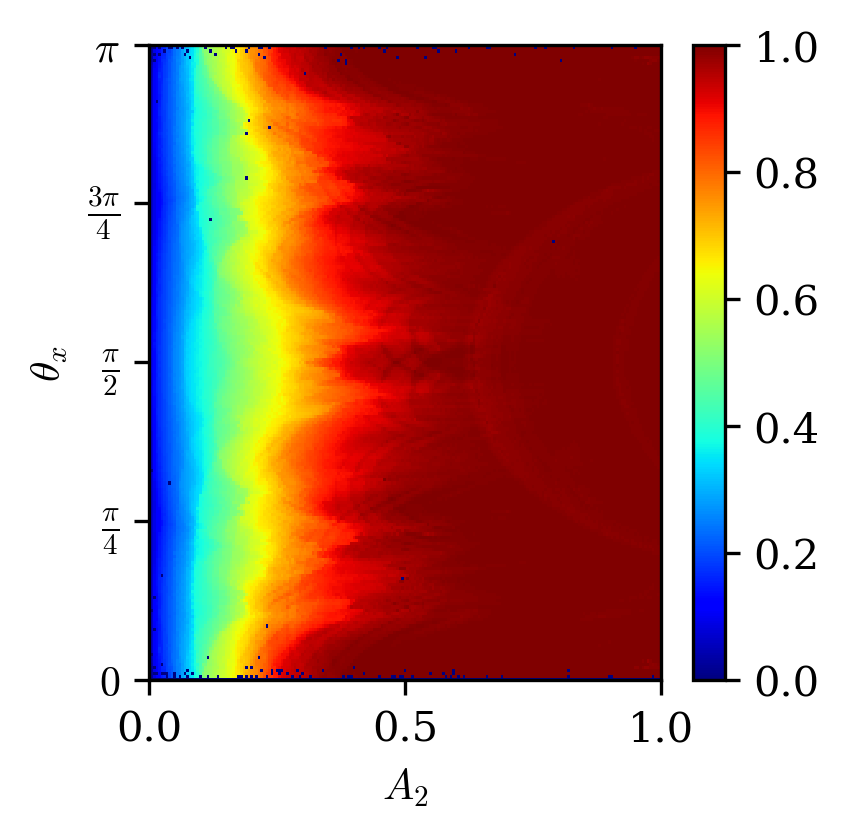}
         \caption{Chaotic}
         \label{fig_chaos}
     \end{subfigure}
     \begin{subfigure}[b]{0.42\textwidth}
         \centering
         \includegraphics[width=\textwidth]{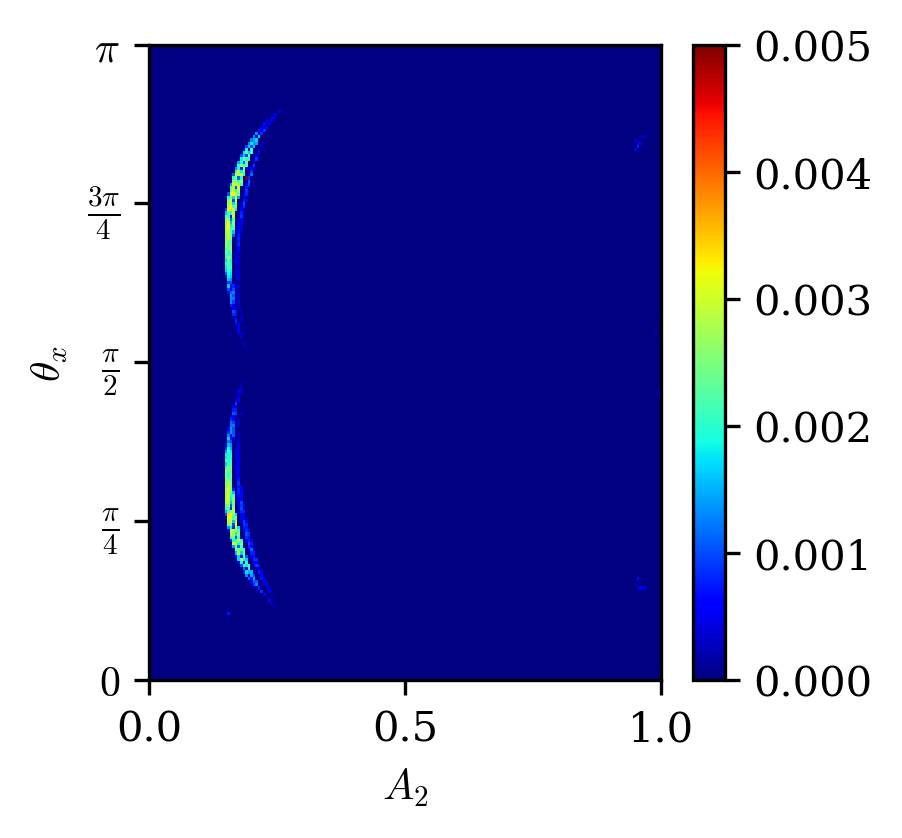}
         \caption{Ballistic}
         \label{fig_ball}
     \end{subfigure}
        \caption{The parameter spaces with the relative area given by the color scale for chaotic and ballistic regimes.}
        \label{fig_pspace}
\end{figure}

We repeat our method for a set of constants sampling the parameter space and compute the relative areas, obtaining Figure \ref{fig_pspace}. \noindent

Figure \ref{fig_pspace} (a) presents a parameter space of the relative chaotic area of the map, where, as expected, with the increase of $A_2$, the chaotic region expands, but this expansion is not linear, saturating for most values of the phase $\theta_x$ around $A_2 \approx 0.5$. It is also interesting to notice that the structures vary depending on the value of $\theta_x$, as displayed by the crests and arches present, as well as the expected symmetry around $\frac{\pi}{2}$ due to the $\sin(\theta_x + xk)$ term.

When it comes to identifying anomalous transport, the routine serves its purpose. In figure \ref{fig_pspace} (b), the region of the phase space with anomalous transport due to the ballistic mode forms two symmetrical arch-like structures, and only in that region, so superdiffusion not only is present but is also very sensitive to a change in these two parameters.

\section{Conclusions}

The presented paper displayed the novel technique of segment and test, and it was possible to identify superdiffusion behavior through the quick identification of islands that display ballistic-like behavior.

Of course, some drawbacks are present; in this case, there is a limit (resolution) to the size of the smallest island detectable, which is determined by the structuring element of the filtering. 

A demonstration was presented on the standard map, a well-known system, where the superdiffusion on the momentum $v$ was identified thanks to ballistic modes, as well as the transition to global chaos. 

The two-wave system was a better use case, as its numerical integration is computationally expensive. In this case, it was also possible to identify superdiffusion due to ballistic-like modes in the parameter space of $\theta_x \times A_2$, where the anomalous transport is restricted only to a thin region. 

Another feature of the method is the possibility to identify major changes in the structure of the phase space, by searching for abrupt changes in the relative areas of some transport regime

Some next steps to improve this routine may be related to better morphological filterings, changes in the decision trees, and of course implementation in different systems.

\section{Acknowledgments}

RLV and AFB thank the Coordenação de Aperfeiçoamento de Pessoal de Nível Superior – Brasil (CAPES) – Finance Code 001 and the Brazilian Federal Agency (CNPq) under Grant Nos. 403120/2021-7, 301019/2019-3. ILC thanks Fundação de Amparo à Pesquisa do Estado de São Paulo FAPESP 2018/03211-6. Centre de Calcul Intensif d’Aix-Marseille is acknowledged for granting access to its high-performance computing resources. Last but not least, we thank members of the Oscillation Control Group at USP and the PIIM laboratory at AMU, for fruitful discussions and insights about the work.

\bibliographystyle{ieeetr}
\bibliography{ref}% Produces the bibliography via BibTeX.

\end{document}